\documentstyle[eqsecnum,floats,prd,aps]{revtex}
\input psfig.sty
\begin{document}
\draft
\newcommand{\be}{\begin{equation}}
\newcommand{\ba}{\begin{eqnarray}}
\newcommand{\ee}{\end{equation}}
\newcommand{\ea}{\end{eqnarray}}
\def\ta{<\! \tau\! >}
\def\la{<\! \lambda\! >}
\def\ar{{r}}

\title{Chaos may make black holes bright}
\author{Janna Levin}
\address{Astronomy Centre, University of Sussex}
\address{Brighton BN1 9QJ}
\address{and}
\address{Center for Particle Astrophysics,
UC Berkeley}
\address{Berkeley, CA 94720-7304}

\maketitle
\widetext
\begin{abstract}
Black holes cannot be seen directly since they
absorb light and emit none, the very quality which earned
them their name.  We suggest that black holes may be seen indirectly
through a chaotic defocusing of light.
A black hole can capture light from a luminous companion in chaotic orbits
before scattering the light in random directions.
To a distant observer, the black hole would appear to light up.
If the companion were a bright radio pulsar, this estimate suggests the
black hole echo could be detectible.

\end{abstract}

\begin{picture}(0,0)
\put(410,210){{ CfPA-98-TH-19}}
\end{picture} \vspace*{-0.15 in}

\twocolumn
\narrowtext

\setcounter{section}{1}

Black holes evade direct detection precisely because they are black.
The existence of black holes hidden behind accretion disks or in the
centers of galaxies have been inferred from astrophysical observations.
Despite these indirect observations, we cannot know for certain that 
the compact objects lurking there are in fact Einstein's black holes.
Any detection which can see in very near to the event horizon would
provide more
incriminating evidence for their existence.
In this {\it Letter} we describe how chaotic scattering of light
in an inner regime around the event horizon
could effectively render the 
black hole bright.
If perturbed, a stochastic region 
develops around the last unstable photon orbit.  
The disturbance could be an orbiting 
companion or the emission of gravitational waves or any asymmetry 
in the evolution.
The black hole can then
trap incident photons in the stochastic region for some time
before throwing off half and absorbing the other 
half,  effectively shrouding it in light.
We estimate the cross-section for this gravitational
defocusing generically and then illustrate with an 
extremal binary black hole spacetime.

Around an isolated black hole 
of mass $M$ and charge $Q$,
light follows the simple orbits 
	\be
	\left (dR\over d\phi\right )^2={R^4\over b^2}-
	R^2\left (1-{2M\over R}+{Q^2\over R^2}\right ) \ \ .
	\label{veff}
	\ee
The motion of massless particles
depends only on the impact parameter $b=L/E$
and not on the energy $E$ and the angular momentum 
$L$ separately.
Given the mass and charge of the black hole, there is a critical 
value of the impact parameter, $b_c$, at which light gets trapped in 
a perfectly circular, unstable orbit.
For smaller impact parameters, light will fall into the black hole
while for larger impact parameters, light escapes.
Just above $b_c$, the light
can come close to the circular orbit executing one or more full rotations 
before
being cast off.  Only black holes are compact enough to bend light by 
more than $\pi$.  The phenomenon of back scattering by a full $\pi$ is
known as the glory \cite{mtw} and was thought to be the weakest way to 
observe black holes.

In the stochastic region around a black hole pair, 
the last unstable orbit becomes the site for chaotic scattering.
The lone periodic orbit is replaced by a glut of periodic orbits.
These proliferating orbits
are packed so densely into phase space that they form a fractal set.
Fractals are a way of maximizing the area while maintaining a bounded 
volume.  Light scatters 
chaotically as it skips from one periodic orbit to another.
The cross-section
for catching light in multiple windings around the hole
is then 
amplified.  As well, the black hole hangs on to the light for longer
and reemits the light more evenly.  A bright light directed onto 
the black hole, say from a pulsar companion, could illuminate
the black hole for a time before the light decayed away and the 
star fell dark again.

\begin{figure}
\centerline{\psfig{file=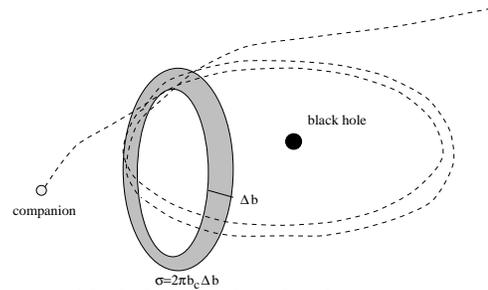,angle=-90,width=2.5in}}
\caption{A black hole with a bright companion captures light around
$R=3M$ for a few windings before the light scatters off.
\label{tar}}  \end{figure} 

The defocusing cross-section can be approximated generally in terms
of the topological features of the chaotic fractal set.
The importance of the general approach is that the scattering
cross-section for any candidate
system can be estimated in this manner.
We assume isotropy which will in fact be broken 
by an actual perturber.  
The perturbation is largest at the point of closest approach and 
is presumably
larger in the orbital plane.
This whole system is also moving 
relative to the observer.
For the purposes of estimating the magnitude of defocusing, 
it is reasonable to assume
isotropy.  In fact, isotropy could be approached for 
complex orbital motions since 
light may no longer be confined to a plane.
The cross-section $\sigma$ is roughly the
geometric area of the annulus 
	\be
	\sigma=2\pi b_c\Delta b
	\ee
as shown in fig.\ \ref{tar}.
As light is shot at the black hole with impact parameter near 
$b_c$, it will travel on nearly periodic orbits for a time before
diverging from these unstable worldlines.  The cross-sectional
thickness $\Delta b $ 
is thus given by the number of periodic orbits 
times the thickness around each orbit in phase
space $\epsilon$.
With $N(n)$ the number of fixed points lying
on periodic orbits which execute
$n$ windings around the black hole,
$\sigma $ can be written as a sum over all winding numbers,
	\be
	\sigma =2\pi b_c^2 \sum_{n=1}^{\infty} N(n)\epsilon(n)
	\  \  .
	\ee
The number of fixed points is given by the 
topological entropy
	\be 
	S=\lim_{n\rightarrow \infty} {1\over n}N(n)\label{entro}
	\ee
so that $N(n)\sim e^{Sn}$
in the limit of long orbits.  
Perturbatively, the deviation is $\delta r/r \sim \lambda $
with $\lambda $ the Lyapunov exponent,
a measure of the instability of the orbit.
The width in phase space is then
$\epsilon(n)\sim b e^{-\lambda n}$
The instability can vary 
from orbit to orbit.  As an approximation we take $\lambda $ to be
the average over all the fixed points.  
Although the Lyapunov exponent is 
a notoriously coordinate dependent quantity, in this context
we are measuring the instability in units of windings around the
black hole and the winding number does not vary from observer to 
observer.
Under the assumption of ergodicity, half of the photons 
fall into the black hole
and half are cast off, so we divide the cross-section in half to find
	\be
	\sigma=\pi b_c^2\sum_{n=1}^{\infty} \left (\exp(S-\lambda)\right )^n
	\ \ .\label{sumr}
	\ee
Performing the sum gives 
	\be
	\sigma =\pi b_c^2{e^{(S- \lambda)}\over 1-e^{(S- \lambda)}}
	\ \ .\label{step}
	\ee
We can recast $\sigma $ in terms of the fractal properties of the set of
periodic orbits.  The fractal
dimension is defined as
	\be
	D=\lim_{\epsilon \rightarrow 0}{\ln N(\epsilon)\over \ln (1/\epsilon)}
	\ \ ,
	\ee
where $N(\epsilon)$ is the number of boxes of size $\epsilon $ needed to 
cover the set.
We can relate $D$ to the entropy by noting that 
$\epsilon\propto e^{-\lambda n}$ for orbits of length $n$,
from which it follows that
		\begin{eqnarray}
	D& =&\lim_{n\rightarrow \infty}{\ln N(\epsilon)\over
	 n\lambda}\nonumber \\
	& =& {S\over \lambda}
	\ \ .\label{dsl}
	\end{eqnarray}
Again, this assumes $\lambda $ is the same across the set or that a 
suitable average will fare well enough.
There is an entire spectrum of 
weighted entropies and dimensions to characterized the fractal 
for which 
similar relationships to (\ref{dsl}) 
have been conjectured \cite{{greb},{ott}}.
Using (\ref{dsl}) in (\ref{step}), we then estimate the cross-section to be 
	\be
	\sigma =\pi b_c^2{e^{- \lambda(1-D)}\over 1-e^{- \lambda(1-D)}}
	\label{area} \  \ .
	\ee

We could have deduced the cross-section directly from the fractal 
property.
The proliferation of periodic orbits form a fractal set which 
fills an area in the oribital plane.  The set, by definition of the
fractal dimensionality, therefore has a thickness.
That thickness is determined 
by the fractal dimension.  According to the usual determination of the
width of a fractal set, 
the number of fixed points
in the direction $dr$ which can be covered by boxes of 
size $\epsilon$ is given by 
$N\sim (\epsilon/b)^{-D}$.
The length of this set is then 
$N\epsilon=b(\epsilon/b)^{1-D}$
with $\epsilon \sim b e^{-\lambda n}$.
Summing and dividing by 2 
we again derive the area
of eqn.\ (\ref{area}).

\begin{figure}
\centerline{\psfig{file=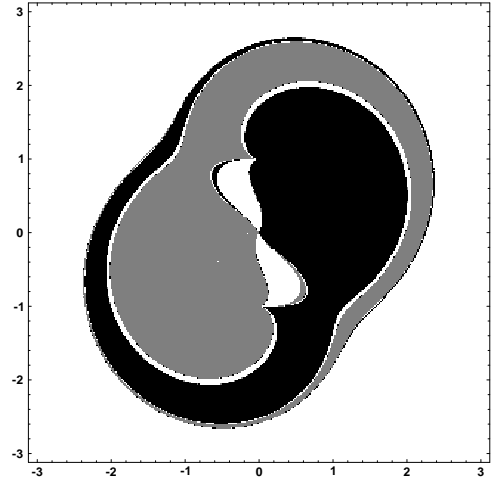,width=2.5in}}
\centerline{\psfig{file=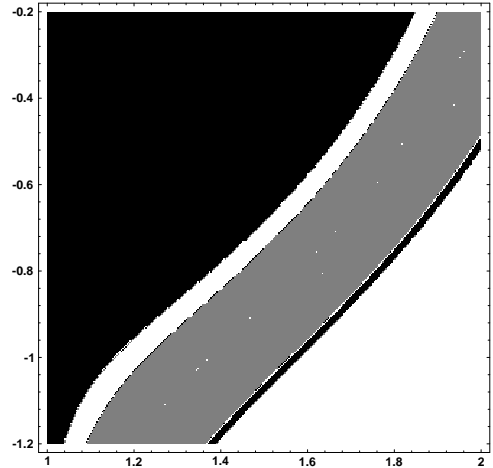,width=2.5in}}
\caption{The color coded fractal basin boundaries
for two extremal black holes with $Q=M=1$.  The initial location 
of the photon is painted black if it sticks to the mass at $(0,1)$,
grey if it sticks to $(0,-1)$, and white if it escapes.
The initial radial velocity is $\dot r=0$ and $\dot \phi >0$ is 
set by eqn.\ (\ref{ham}).
The error tolerance was $\sim 10^{-8}$.
\label{fbb}}  \end{figure} 

We can conservatively evaluate $\sigma $ by using the unperturbed 
values.
The dimension of the boundary is zero in the unperturbed, nonchaotic
system.
To estimate $\lambda $, we vary eqn.\ (\ref{veff}) around 
the unstable circular orbit.  
For a chargeless Schwarzschild black hole of mass $M$, the last
unstable circular orbit lies at $R=3M$ with $b_c=3\sqrt{3}M$.  
The perturbed radial motion grows as $\delta R/R\sim e^{-\phi}$, or 
in units of $\phi=2\pi n$, the exponent is $\lambda =2\pi$.
We then estimate $\sigma\sim {\cal O}(10^{-3})\pi b_c^2$.
For an extremal
black hole ($Q=M$), the last
unstable circular orbit
occurs at $R=2M$ with $b_c=4M$.
At second order we find 
$\delta R/R\simeq e^{- \phi/\sqrt{2}}$.
Measuring $\phi $ in terms of the winding number, $\phi\sim 2\pi n$,
we read off $\lambda\simeq \sqrt{2}\pi $.
For an extremal black hole, the cross-section is larger with
$\sigma \sim {\cal O}(10^{-2})\pi b_c^2 $; that is,
a hundreth the capture cross-section.

Unlike other more conventional estimates of the width of a stochastic
layer, this estimate requires knowledge of only a few simple
properties of the system and does not require a complicated 
examination of the dynamical equations.
Unlike other more conventional estimates, there are inherent 
shortcomings.
As with the glory calculations \cite{mtw}, the effect is dominated by 
the trajectories which wind around the black hole the fewest number
of times.
These trajectories may not model the fractal set of underlying 
periodic orbits as well as those which execute many windings and spend
the most time chaotically scattering off the set.  In other words, the
sum in eqn. (\ref{sumr}) is dominated by the first few $n$ while eqn. 
(\ref{entro}) is a large $n$ limit.
Given this caveat, we can see how well the approximation fares in a
given dynamical system.
To explicitly illustrate, we consider scattering around two 
extremal black holes.
A pair of black holes with equal charge and mass 
are able to coexist in a static configuration with the electrostatic
repulsion caused by their charge just balancing the gravitational
attraction of their masses.
While the resultant Majumdar-Papapetrou spacetime \cite{{maj},{pap}}
is static, the geodesic flows around the pair of 
black holes are known to be chaotic \cite{{bhs},{carl}}.
We isolate fractal basin boundaries \cite{{carl},{us},{mixm},{gc},{ym}}
for massless particles
as has already been done for massive particles \cite{carl}.

The Lagrangian for motion in this space can be written in 
isotropic coordinates as
	\be
	{\cal L}=-{1\over 2}U^{-2}\dot t^2+
	{1\over 2}U^2\left(\dot \ar^2+\ar^2\dot \Omega^2\right )
	\ee
with $\dot\Omega^2=\sin^2\theta \dot\theta^2+\dot\phi^2$ and 
an overdot denotes differentiation with respect to an affine parameter.
Schwarzschild coordinates are recovered with $R=Ur$.
The metric components are determined by
	\be
	U=1+{M\over r_1}+{M\over r_2}
	\ \ ,
	\ee
where $r_1$ is the coordinate distance from one black hole 
and $r_2$ is the coordinate distance from the other.
The event horizons occur at $r_1=0$ and at $r_2=0$.
We place a black hole with mass $M $ 
and
charge $Q=M$
at $(r,\theta,\phi)=(M,\pi/2,\pi/2)$
and an identical companion at $(M,\pi/2,-\pi/2)$ so that 
	\ba
	r_1^2&=&r^2-2Mr\sin\phi+M^2 \nonumber \\
	r_2^2&=&r^2+2Mr\sin\phi+M^2
	\ \ .
	\ea
The geodesic motion is found by evolving the conjugate momenta,
$\Pi_q={\partial {\cal L}/\partial \dot q}$.  Two coordinates are 
automatically conserved with 
	\ba
	\Pi_t &=& {\dot t\over U^2}=E\nonumber \\
	\Pi_\theta &=& r^2U^2\dot \theta=L_\theta \ \ .
	\ea
while two are dynamical 
	\ba
	\Pi_\phi &=& r^2U^2\dot \phi \nonumber \\
	\Pi_r &=& U^2 \dot r \ \ 
	\ea
and evolve according to the equations 
$\dot \Pi_q={\partial {\cal L}/\partial q}$.
There is an additional constraint equation, equivalent to the conservation 
of energy obtained by setting
${\cal L}=0$ 
so that the photons travel
along null geodesics:
	\be
	\dot r^2+r^2\dot \phi^2=E^2\ \ .
	\label{ham}
	\ee
We have assumed that $L_\theta =0 $ and so consider motion in the 
plane defined by the binary system ($\theta =\pi/2$).

We numerically evolve the goedesic equations, ensuring that the 
energy is conserved according to eqn.\ (\ref{ham}). 
In isotropic coordinates, the last unstable
circular orbit around an isolated black hole occurs at 
$r=M $.  So, in the black hole pair, we expect to see chaos around 
$r_1=M$ and $r_2=M $, which is precisely what we find as illustrated by 
the
fractal basin boundaries of 
fig.\ \ref{fbb}
\cite{{carl},{us}}.  We look at an initial slice
through the plane of the binary system.  We submerge 
the black holes and their
surrounding area in a bath of light, a photon at 
each location in space with $\dot r=0$ initially and 
$\dot\phi>0 $ set by eqn.\ (\ref{ham}).  
We color code the initial location black  
if the photon which originated there fell into the hole
at $r_1=0$, grey if it fell into $r_2=0$, and white if
it escaped.
The result is the mixed, fractal basin
boundary as shown in fig.\ \ref{fbb}.  The lower panel focuses in on 
one region to show the repetition of the fractal structure.
The dimension of this boundary 
is $1+D\simeq 1.1$.

If we take $\Delta b$ to be roughly the size of the white strip in the 
lower panel of fig.\ \ref{fbb}, we would estimate $\delta r/M\sim 0.02 $.
Of course, the meaning of this coordinate thickness is ambiguous.
Still it is reassuring that it is comparable
in magnitude to the value we would have guessed from our generic approximation
using the unperturbed $\lambda $ and the measured
value of $D\sim 0.1$, which gives
$\Delta b/b_c \sim 0.02$, although the extreme agreement is certainly
fortuitous.
The cross section is again about $\sigma \sim {\cal O}(10^{-2})\pi b_c^2 $.

It is not immediately obvious if the approximation underestimates or
overestimates the corss-section.
Some elements are underestimated.
For instance,
there are additional contributions to the defocused light 
from the inner regions of the binary system which are not included in the
estimate of eqn.\ (\ref{area}).
Also,
any realistic astrophysical black hole with a companion will only be
more chaotic than this crisp example.  The metric will not be static
and it is likely that the evolution of the spacetime itself will
be chaotic.  The companion needed to provide the luminous source may 
well be on an unpredictable trajectory contributing to the probability
of defocusing.
On the other hand, the calculation was restricted to the orbital plane
where the effect would be largest and the relative orientation of the
observer may decrease the signal.

\begin{figure}
\centerline{\psfig{file=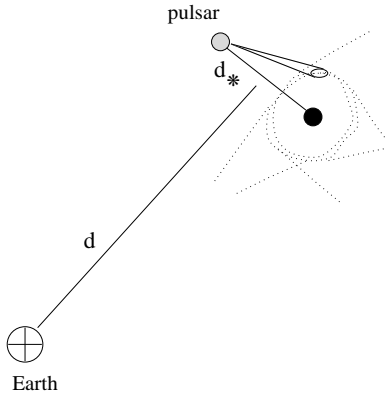,angle=-90,width=2.0in}}
\caption{A black hole, pulsar, Earth configuration.
\label{compan}}  \end{figure} 

We can use the estimate of the cross-section from the 
Majumdar-Papapetrou spacetime to give rough predictions for the
observability of the
effect.
To a distant observer, a black hole with a pulsar companion
will
appear to radiate with a 
luminosity
	\be
	L_{\rm BH}\simeq L_*\left (\sigma\over \sigma_{beam}\right )
	\left (\tau_*\over \tau \right )
	\ \ 
	\ee
where $L_*$ is the observed
luminosity of the star, the binary lies at a distance 
$d$ from the Earth,
and $d_*$ is the relative
star and black hole separation as shown in Fig.\ \ref{compan}.
The cross-section of the beam is roughly $\pi d_*^2\sin^2(\theta_{beam})$
where $\theta_{beam}$ is the half-angle subtended by the pulse.
We further assume that the time it takes the beam to sweep over the 
black hole $\tau_*$ is comparable to the characteristic
time $\tau $ for the captured light to decay off the black hole.
We take the companion to be at 
a distance of $d_*\sim 10^n M_{\rm BH}$.
Using $\sigma \sim {\cal O}(10^{-3})\pi 27 M^2$ and assuming
$\theta_{beam}\sim 5^o$ we estimate
	\be
	L_{BH} >  3.6 \times 10^{-2n+1} \ L_*
	\ \ .
	\ee
Notice that this is the most pessimistic estimate.  The black hole
luminosity is suppressed relative to the incident luminosity by
$\sigma/\sigma_{beam}$.  Since $\sigma_{beam}$ is much larger than the
scattering cross-section, the returned radiation looks small.
However, the incident radiation can accumulate as the pulse returns again and
again feuling a larger
returned radiation.
In fact, this is how pulsars are observed here on Earth.  The
entire beam luminosity is collected as the signal sweeps across the
telescope.  Pursuing this rough most conservative estimate
for a pulsar at a distance $d_*=100 M$, 
a signal $10^{-4}$ times fainter 
would be seen just behind the original pulse.
The luminosity is also transient, decaying in a timescale
related to the instability of the orbits.  
(For a solar mass black hole $\tau\sim 2GM_\odot/c^3\sim 10^{-5}-10^{-6}$ sec.)
Even if the beam pointed away from the Earth, as
it swept over the black hole it would feed the stochastic layer and 
we would see a faint echo of the unseen pulsar from the diffuse light 
defocused off the black hole.
The gravitational Doppler shift 
will also separate the frequency of the 
echo from that of the original pulse for nonstatic systems.
As a last point, superradiance of scattered light from the ergosphere
of a rotating black hole could also be significant.

A more realistic calculation will be challenging as is reflected
by the imfamous full relativistic binary problem.  What is clear 
is that 
as the companion gets closer, chaos will be more important and the
signal will get brighter, but the 
lifetime of the binary will also be shorter.
The last stages of inspiral may be characterized by the defocused
echo in coordination with the gravity wave signal expected.

Most black hole systems currently accessible to observation involve
the accretion of material from a luminous companion and the defousing
effect would be
completely obscured.
Only in the most minimal binary pairs will the chaotic scattering lead
to a
visible glow around the black hole
such as black hole/neutron star pairs and black hole/puslar 
pairs in particular.  Since these are amoung the systems the future gravity
wave experiments hope to discover, this could offer a valuable 
electromagnetic observational counterpart to any gravity wave
detection.
The gravity wave experiments hope to detect quite distant coalescing
binaries.  It is unlikely that a pulsar would ever be visible at such
large distances.  
Nonetheless, during the last stages of inspiral tidal stresses
will undoubtedly heat up the companion providing a brighter
electromagnetic signal to observe.  The details of such a scenario are
far from clear but the possibilities are worth investigating.

While we have been promoting chaotic defocusing as a means to view
the inner orbits around a black hole, there are other observable 
consequences of the chaotic flows.
For instance gravity-waves are a natural and inevitable source 
for the perturbations \cite{{cal},{moeckel}} and the   
implications for the direct detection of gravity-waves from the 
last stages of inspiral in a compact binary
will certainly be significant.
More immediately observable, the disrupted motions of 
an accretion disk around black hole
candidates
could lead to an indirect detection of gravity-waves. 
Whether in radio waves or gravity waves, chaos may in fact make black holes
bright.

\vskip 15truept

I am particularly grateful to Neil Cornish for
numerous critical discussions.  Many of these ideas grew out of our
related collaborations.
I also thank John Barrow,
Pedro Ferreira, John Hibbard, Andrew Jaffe and Mike Eracleous 
for their interest and valuable comments.

\end{document}